\documentclass[11pt]{article}
\usepackage{graphicx}  
\usepackage{amssymb,latexsym}
\usepackage[mathscr]{eucal}
\usepackage{url}

\newcommand{\bea}{\begin{eqnarray}}
\newcommand{\bean}{\begin{eqnarray*}}
\newcommand{\eean}{\end{eqnarray*}}
\newcommand{\beal}[1]{\begin{eqnarray}\label{#1}}
\newcommand{\eea}{\end{eqnarray}}

\newcommand{\be}{\begin{equation}}

\newcommand{\ee}{\end{equation}}

\begin{document}

\title{Penrose's circles in the CMB and a test of inflation}

\author{Paul Tod\\Mathematical Institute\\St Giles', Oxford OX1 3LB}

\maketitle
\begin{abstract}
We present a calculation of the angular size of the circles in the CMB predicted by Penrose on the basis of his Conformal Cyclic Cosmology. 
If these circles are detected, the existence of an upper limit on their angular radius would provide a challenge for inflation.
\end{abstract}


\medskip

Penrose has described a radical new view of the universe, Conformal Cyclic Cosmology, in talks and articles over the past five or six years \cite{rp0}, \cite{rp00}, with the fullest account in the book \cite{rp1}.
 At an early stage, he remarked that a prediction of CCC was that there should be circular structures observable in the CMB arising from events late in conformal time in the previous aeon. 
Late in conformal time, the content of the universe according to CCC is radiation and supermassive black holes, and the only significant events are black-hole mergers which give rise to sharp bursts of 
gravitational radiation. This radiation travels from one aeon to the next and perturbs the matter distribution early in the next aeon, which in turn produces circular perturbations in the observed CMB. In two recent articles \cite{rp2}, \cite{rp3}, he and 
Gurzadyan claim that these circles can be observed in the CMB as circles with significantly lower variance in the temperature. This claim is controversial and a number of authors have disagreed 
with the statistical significance of the findings of Penrose and Gurzadyan \cite{we}, \cite{msz}, \cite{h}, \cite{ew2}.

In this article, we do not propose to enter the debate about whether the circles have in fact been observed, but rather 
to suppose that Penrose's picture of CCC is correct in order to follow the consequences further. Subject to simple 
assumptions which are made explicit, we present a short calculation of 
the angular sizes of circles of the kind hypothesised by Penrose and show that there is an upper limit to their angular radius of around 21 degrees. We then remark that 
if, in a CCC model, there is a period of inflation, as usually understood then the upper limit goes away and the circles can have any angular radius (i.e. up to 90 degrees). Thus if the circles are 
what Penrose and Gurzadyan claim them to be, or if later observations are acknowledged to detect circles, and if there is an upper limit on their radius  
similar to that suggested here then this would constitute an observational challege to the theory of inflation as usually understood. 
(A calculation of the distribution of angular sizes of the proposed circles has been given in \cite{nwe}, 
with different aims, though the effects of inflation are considered there.)

For the calculation, assume the FRW form of the metric:
\[
ds^2=dt^2-(R(t))^2(dr^2+f_k^2(r)(d\theta^2+\sin^2\theta d\phi^2)),
\]
where as usual $f_k=\sin r, r, \sinh r$ according as $k=1,0,-1$. CCC is particularly simple to describe with the FRW metric: the scale factor $R(t)$ runs from zero at the Big Bang of one aeon through to infinity 
at the future infinity of that aeon, which is identified as the Big Bang of the next aeon. Introduce conformal time $\tau$ in terms of proper time $t$ by $dt=Rd\tau$ and then the FRW metric can be written
\be\label{g1}ds^2=(R(t))^2(d\tau^2-dr^2-f_k^2(r)(d\theta^2+\sin^2\theta d\phi^2))=(R(t))^2d\tilde{s}^2,\ee
and then the conformal metric $d\tilde{s}^2$ is cyclic: it extends smoothly through each Big Bang and future infinity. These are separated by infinite intervals of proper time but finite intervals of conformal time.

As is conventional, we assume the matter content of the universe to be a mixture of 
dust and radiation together with a positive cosmological constant $\Lambda$. The Einstein equations reduce to the Friedman equation:
\be\label{f1}\dot{R}^2=-k+\kappa (\rho_M+\rho_\gamma) R^2+\frac13\Lambda R^2,\ee
where $\kappa=8\pi G/3$, $\rho_M$ and $\rho_\gamma$ are the matter and radiation densities, and the overdot is $d/dt$ with proper-time $t$, together with the perfect fluid conservation equation 
\[3\dot{R}/R=-\dot{\rho}/(\rho+p).\]
The conservation equation for the two fluids can be solved to give
\[\rho_M=AR^{-3},\;\;\rho_\gamma=BR^{-4},\]
in terms of integration constants $A,B$. For simplicity assume $k=0$, though data given in, for example, \cite{DM} indicates that this term in the Friedman equation is in any case very small.

Now the Friedman equation is just
\[\dot{R}^2=\kappa(AR^{-1}+ B R^{-2})+\frac13\Lambda R^2.\]
For the scale factor at the present time, write $R=R_0$ and introduce constant parameters $a,b$  by
\[\Omega_M/\Omega_\Lambda=a,\;\;\Omega_\gamma/\Omega_\Lambda=b,\]
at the present time, with the conventional meanings for $\Omega_M, \Omega_\Lambda$ and $\Omega_\gamma  $. This translates to 
\[
\kappa \rho_M/(\frac13\Lambda)=a,\;\;\kappa \rho_\gamma/(\frac13\Lambda)=b,
\]
at the present time, and hence 
\[\kappa A/(\frac13\Lambda R_0^3)=a,\;\;\kappa B/(\frac13\Lambda R_0^4)=b.\]
One may find values for $(a,b)$ in the literature including $(a,b)=(.33, 10\times 10^{-5})$ in \cite{DM} or $(.35, 6.5\times 10^{-5})$ 
in \cite{pdg}, while \cite{nwe} uses $(.15,5.6\times 10^{-5})$.

Eliminate $A,B$, introduce $S=R/R_0$ and rationalise (\ref{f1}) as:
\be\label{f2}
S^2\dot{S}^2=\frac13\Lambda(b+aS+S^4).
\ee
We shall be interested in conformal time $\tau$ rather than proper time $t$, so $dt=Rd\tau$ and for a single aeon we may choose the origins to coincide. Then (\ref{f2}) reduces to the integral
\be\label{f3}
\int_0^S\frac{dS}{(b+aS+S^4)^{1/2}}=\tau R_0\sqrt{\frac{\Lambda}{3}},
\ee
where $\tau=0=t$ at $S=0$.

There are three values of $\tau$ of interest: first, that corresponding to now, say $\tau_0$, obtained when the upper limit in the integral is at $S=1$; and second that corresponding to the remote future, 
say $\tau_\infty$, obtained when 
$S=\infty$. The constants on the right in (\ref{f3}) cancel from the ratio $q$ of these, which is therefore
\[
q:=\tau_0/\tau_\infty=\left(\int_0^1(b+aS+S^4)^{-1/2}dS\right)\left(  \int_0^\infty(b+aS+S^4)^{-1/2}dS\right)^{-1}.
\]
Evaluating this for different choices of $(a,b)$, we obtain $q=0.76$ following values given in \cite{DM} or \cite{pdg}, or 0.81 for values given in \cite{nwe}. (Note that this 
means that, in conformal time, the universe is now at least three-quarters of its total age.)
 
There is a very weak dependence on 
$(a,b)$ of the ratio $(\tau_\infty-\tau_0)/\tau_\infty$, which is the fraction of conformal time remaining, because in the range $1\leq S\leq\infty$ the terms 
$b$ and $aS$ in the integrand are swamped by the term $S^4$, and the integral is well-approximated by
\[\int_1^\infty(S^4)^{-1/2}dS=1.\]
Part of Penrose's theory of CCC is that all matter eventually becomes massless, \cite{rp1}. The weak dependence on $a,b$ just observed implies that the precise epoch at which this transition occurs, 
and the details of the process, do not affect the calculation being done here.

\medskip

We also need the ratio $\tau_{LS}/\tau_\infty$ where $\tau_{LS}$, the third value of $\tau$ we need, corresponds to the time of the last-scattering surface. According to \cite{nasa}, 
the redshift at decoupling  is 1089 so the scale 
factor then was $R_0/1090$ and $S$ was $S_{LS}=1/1090$. We may use $(a,b)$ from \cite{DM} to find $\tau_{LS}/\tau_\infty=.015$, with similarly small values using \cite{nwe} or \cite{pdg}. This is small 
compared to $q$, a fact equivalent to the `horizon problem', 
but the calculation leading to the value of $\tau_{LS}$ changes 
completely in the presence of inflation, as normally pictured.

\medskip

For the geometry of the circles, note from (\ref{g1}) that the equation of the light cone of the origin is $\tau\pm r=$constant, where plus is for the past light cone and minus 
for the future one. Thus the light cone of the origin at conformal time $\tau_1$ meets the hypersurface at conformal time $\tau_2$ in a sphere of coordinate radius $|\tau_1-\tau_2|$. 
Suppose some emission event in the previous aeon, due to a late-$\tau$ black hole merger, occurred at $\tau_e$, and sent out a burst of gravitational radiation supported close to its 
future light cone. It is the intersection of this cone from the previous aeon with the last-scattering surface in our aeon which produces the circular structures predicted by Penrose. 
We need to assume something about the previous aeon. CCC is not developed to the point where one can calculate the parameters, for example $A$ and $B$ above, determining one aeon from their values in the previous aeon. 
In the absence of such a theory, the simplest assumption to make is that 
our aeon and the preceding one were similar to the extent of having similar evolutions and similar values for the constants $A,B$ (choosing $\Lambda$ the same from aeon to aeon 
is essentially a gauge choice); then $a,b$ and the range in $\tau$, say 
$0\leq\tau\leq\tau_\infty$ are similar. 
%
%

The future light cone from the black hole merger meets the last-scattering surface in a sphere of coordinate radius 
$\tau_\infty-\tau_e+\tau_{LS}$, where, as before, $ \tau_{LS}$ is the 
conformal time between the bang and the last-scattering surface. Our past light cone, starting from $\tau_0$, meets the last-scattering surface in a sphere of 
radius $\tau_0-\tau_{LS}$. 
These two spheres, if they meet, will meet in a circle, which is Penrose's predicted circular structure, and the largest such circle has angular radius $\theta$ with
\[\sin\theta=(\tau_{LS}+\tau_\infty-\tau_e)/(\tau_0-\tau_{LS})\]
provided this ratio is less than one. We shall assume that $\tau_e\geq\tau_0$ i.e. that the presumed super-massive black 
hole mergers generating the circles, which according to CCC occur late in $\tau$-time when the universe contains just radiation and black holes, happen 
at some stage in the previous aeon which has not yet been reached in this one. Then the maximum value of 
$\sin\theta$ can be calculated as 0.35 following \cite{DM}, 0.36 following \cite{pdg}, or 0.26 following \cite{nwe}, corresponding to maximum angular radii of 20.5 degrees, 21 degrees or 15 degrees respectively. Evidently, later 
emission will give smaller circles, so these are upper limits under the assumptions made. (In \cite{rp1} a maximum angular radius of 30 degrees was suggested based on the ratio $q:=\tau_0/\tau_\infty$ 
being about 2/3, a figure that we are seeking to refine here.)

Now let us repeat the calculation, still within CCC but assuming a period of inflation as usually conceived. In terms of conformal time, one can think of inflation as increasing 
$\tau_{LS}$, the conformal time between the Big Bang and last scattering, to a larger value say $\tilde{\tau}_{LS}$, while not affecting the conformal time intervals $\tau_0-\tau_{LS}$ between last-scattering 
and now, or $\tau_\infty-\tau_0$ between now and infinity (though of course $\tau_0, \tau_\infty$ will increase, say to $\tilde{\tau}_0, \tilde{\tau}_\infty$ respectively). To solve the horizon problem by 
inflation, one needs the sphere arising as the intersection of the  last-scattering surface with our past light cone, which has coordinate radius $\tau_0-\tau_{LS}$, to lie within 
the sphere arising as the intersection of the  last-scattering surfacee with the future light cone of a point on the Big Bang, which has coordinate radius $\tilde\tau_{LS}$. This means that one needs
\[ \tilde\tau_{LS} \gtrsim(\tilde{\tau}_0-\tilde{\tau}_{LS})=(\tau_0-\tau_{LS}).   \]
With what we found above for $\tau_0/\tau_\infty$, this corresponds to $\tilde\tau_{LS}/\tilde{\tau}_\infty\gtrsim 0.4$. This is to be contrasted with the previous calculation which gave $\tau_{LS}/\tau_\infty=.015$ -- 
the role of inflation is to increase this ratio by a factor of about 25 (or much more, depending on the model). Now the quantity of interest for the angular radius of circles is the ratio of conformal time intervals 
from emission to last-scattering and last-scattering to now i.e. 
\[(\tilde{\tau}_{LS}+\tau_\infty-\tau_e)/(\tau_0-\tau_{LS}).\]
If we take the emitting event to occur with $\tau_e=\tau_0$, and substitute the smallest allowed value of $\tilde\tau_{LS}$ this becomes
\[1+(\tau_\infty-\tau_0)/(\tau_0-\tau_{LS}),\]
(larger values of $\tilde\tau_{LS}$ increase this ratio). Evidently this is always greater than one, and in fact it ranges from 1.33 following data in \cite{DM} to 1.24 following data in \cite{nwe}. Since it 
is greater than one, the sphere on the last-scattering surface from the light-cone of emission is larger than the sphere from the light-cone of the observer, 
and so the spheres can intersect along a great circle on the sphere from the observer. This implies the existence of a circular feature along a great circle in the sky. 
Thus, with inflation, the circles in the CMB can have any angular radius (i.e. up to the maximum of 90 degrees), while assuming there is no inflation we found an upper limit of around 
20 degrees. In this context it is worth noticing that the largest circles found in the study reported in \cite{rp2} and \cite{rp3} have angular radius about 16 degrees, \cite{rpn}.

\subsection*{Acknowledgements}
I am grateful to Sir Roger Penrose for many helpful discussions, to the Albert Einstein Institute, Golm, for hospitality while this work was in progress, and to Edward Wilson-Ewing for a helpful e-mail correspondence.

\end{document}